\documentclass[12pt,a4paper]{article}
\usepackage{amssymb}
\def\pa{\partial}

\newcommand{\initiate}{\setcounter{equation}{0}}

\newcommand\be{\begin{equation} }
\newcommand\bea{\begin{eqnarray}}
\newcommand\ee{\end{equation}}
\newcommand\eea{\end{eqnarray}}
\def\ncr{\nonumber\\ }
\def\slash{{\rlap /}}

\def\MBVR{Maja Buri\'c\footnote{E-mail: majab@phy.bg.ac.yu}
\ and Voja Radovanovi\'c\footnote{E-mail: rvoja@ff.bg.ac.yu}\\
{\it Faculty of Physics, P.O. Box 368, 11001 Belgrade, Serbia and
Montenegro}}

\def\endtitle{\par\end{quotation}\vskip3.5in minus2.3in\newpage}

\def\a{\alpha}       \def\b{\beta}
         \def\d{\delta}
\def\e{\epsilon}     
\def\g{\gamma}       
        
\def\k{\kappa}       \def\l{\lambda}
\def\m{\mu}          \def\n{\nu}
       
       \def\r{\rho}
\def\s{\sigma}

         \def\G{\Gamma}

      \def\cl{{\cal L}}

\begin{document}

\title{On divergent 3-vertices in noncommutative $SU(2)$ gauge theory}

\author{\MBVR}

\date{}

\maketitle

\abstract{We analyze divergencies in  2-point and  3-point
functions for noncommutative $\theta$-expanded $SU(2)$-gauge
theory with massless fermions. We show that, after field
redefinition and renormalization of couplings, one  divergent term
remains.}

\vfill \noindent  \eject
\parskip 4pt plus2pt minus2pt

\initiate \section{Introduction and notation}

One of the main motives to study  noncommutativity of the spatial
coordinates is a belief that it will provide a mechanism for
regularization of ultraviolet divergencies. There are various
results which support this idea~\cite{cho,fuzzy}. However, in the
field-theoretic models on  noncommutative Minkowski space,
obstructions to renormalizability appear, as for example the UV/IR
mixing~\cite{UV/IR}. Recently some theories were
proposed~\cite{gw,sla} where renormalizability was restored by
modifications of the original lagrangian.

One should note that the representation of noncommutative fields
plays an important role in  aforementioned considerations. Indeed,
not every representation allows even a definition of the cyclic
trace and thus of the action. In this letter we study
noncommutative gauge theories in the so-called $\theta$-expanded
representation or approach, as given by~\cite{th-exp,th-exp1}.
There are various results regarding renormalizability in this
approach; they could roughly be summarized as `almost
renormalizability'. This means that, in all examples which were
considered, only one divergency in the effective action, the
four-fermion vertex, remained after a generalized renormalization
procedure. This was our motivation for a further and more detailed
analysis of the structure of divergencies; a part concerning the
$SU(2)$ gauge theory is presented here.

To start with, we recall necessary definitions and give a brief
review of the results relevant for our analysis. We discuss  gauge
and matter fields on the noncommutative Minkowski space defined by
the commutation relation between the coordinates $\hat x^\m$ ($\m
= 0,1,2,3$): $$ [\hat x^\mu ,\hat x^\nu ] = i\theta^{\mu\nu} ={\rm
const}. $$ The fields in this case  can be represented in the
space of functions on ${\bf R}^4$; the multiplication is given by
the Moyal product: \be \label{moyal} \phi (x)\star \chi (x) =
     e^{\frac{i}{2}\,\theta^{\m\n}{\pa\over \pa x^\m}{\pa\over \pa
     y^\n}}\phi (x)\chi (y)|_{y\to x}\ .\ee
The basic idea of the approach is that, in order to represent
arbitrary gauge symmetries, one has to enlarge the corresponding
algebra. The gauge fields thus take values in the enveloping
algebra of the given Lie group; the vector potential $\hat A_\r$
can be expanded in the basis of the symmetrized products of
generators $T^a$, $:T^ aT^b\dots :\,$. An important fact is that
the coefficients in this expansion are not independent fields;
they are derivatives of the commutative gauge potential. Moreover,
the expansion (known as the Seiberg-Witten (SW)  expansion,
\cite{sw}) coincides with the expansion in the parameter $\theta
^{\m\n}$. To first order in $\theta^{\m\n}$ it reads: \be
\begin{array}{l}
\hat A_\r(x) =A_\r(x) -\frac 14 \,\theta ^{\m\n}\left\{ A_\m(x),
\pa _\n A_\r(x) +F_{\n \r}(x)\right\}
+\dots\label{expansion}\\[6pt]
\hat\psi (x) = \psi (x)- \frac 12 \, \theta ^{\m\n} A_\m(x)\pa
_\n\psi (x) +\frac {i}{4}\, \theta ^{\m\n}A_\m(x)A_\n(x)\psi
(x)+\dots .
\end{array}                               \label{SW}
\ee We see that the noncommutative field expressed in terms of its
commutative counterpart is  nonlocal, as (\ref{SW}) has infinitely
many terms. The SW expansion  is not given uniquely, and this is
an important fact which we will use later.

The (\ref{SW}) are general expressions, valid for any gauge group.
Here we restrict our discussion to the $SU(2)$ gauge theory with
fermions in the fundamental representation; for the reasons which
we explain shortly the fermions are massless. Inserting (\ref{SW})
in the noncommutative action, in linear order we obtain: \be S=S_0
+S_{1,A}+S_{1,\psi}. \label{S} \ee In the case of $SU(2)$ and
massless fermions (\ref{S}) reads \bea\label{l0} S _{0 }\ &=&\int
d^4x\,\left(\bar\psi \big( i\g ^\m D_\m\big)\psi -{1\over
4}F^{\m\n a}F_{\m\n}^a\right) ,\\ S_{1,A} &=& 0 ,\ncr
\label{l1psi}S_{1,\psi}      &=&\frac 12\,\theta ^{\r\s}\int
d^4x\,\left( -i\bar\psi\g ^\m F_{\m\r}D_\s\psi -\frac i2\,
\bar\psi F_{\r\s}\g ^\m D_\m\psi \right)
\\ &=&
-\frac 18\,\theta^{\r\s}\Delta^{\m\n\a}_{\r\s\b}\int d^4x\, \bar
\psi F_{\m\n}\g^\b (i\pa_\a+A_\a)\psi , \nonumber \eea where $
\Delta ^{\a\b\g}_{\s\r\m} =\d ^\a_\s\d^\b_\r\d^\g_\m -\d
^\a_\r\d^\b_\s\d^\g_\m +({\rm cyclic\ }\a \b \g  )=-\epsilon
^{\a\b\g\l}\epsilon _{\s\r\m\l}\ .$ In the general case the linear
bosonic term $S_{1,A}$ does not vanish; it depends on the
symmetric symbols $d_{abc}$ of the representation of the gauge
potential~\cite{th-exp1}. The first order corrections $S_{1,A} $
and $S_{1,\psi}$ can be treated as new couplings which describe
the effects of noncommutativity. In all orders, the correction
terms are invariant to  the commutative gauge transformations;
they, however, explicitly depend on the choice of representation.
One sees immediately that the action (\ref{l0})
 has a smooth limit $\theta\to 0$; in this limit it
reduces to the ordinary gauge theory. The limit is physically very
important as $\theta$ is small, of the order of magnitude of
$l_{{\rm Planck}}^2$. Of course, the existence of a smooth limit
could be considered as a drawback too, because in many
noncommutative models the commutative limit is singular, for
example in ordinary quantum mechanics.

The fact that the `coupling constant' $\theta^{\m\n}$ is
dimensionfull implies the apparent non-renormalizability of the
theory, unless there is some additional symmetry. The idea that
nonuniqueness of the expansion (\ref{SW}) might play the role of a
symmetry appeared first in~\cite{u1}. They found that the
divergencies in the photon propagator in noncommutative
electrodynamics can be, in  $\theta$-linear order, absorbed in a
redefinition of fields; moreover, that such redefinition can be
generalized to all orders in $\theta$. Let us formulate this more
precisely. Write the SW expansion as \be \label{SWexpansion}\hat
A_\m=\sum A_\m^{(n)} ,\quad \hat \psi =\sum \psi ^{(n)}\ ,\ee
where $A_\m^{(n)}$, $ \psi ^{(n)}$ denote the terms of the $n$-th
order in $\theta$. Then  the transformation \be\label{red}
{A_\m^{(n)}}\to A_\m^{(n)} + {\bf A}_\m^{(n)},\quad
{\psi^{(n)}}\to \psi ^{(n)} + {\bf \Psi}^{(n)}\ ,\ee does not
change the noncommutative fields $\hat A_\m$, $\hat\psi$ if ${\bf
A}_\m^{(n)}$, ${\bf \Psi}^{(n)}$ are gauge covariant expressions
containing exactly $n$ factors  $\theta$ \cite{u1,a}. The change
of the action induced by the shift (\ref{red}) is \be \label{DS}
\begin{array}{l}
\Delta
S^{(n,A)} =\int d^4x\, (D_\n F^{\m\n}){\bf A}_\m^{(n)}\ ,\\[6pt]
\Delta S^{(n,\psi )} =\int d^4x\, \Big( \bar\psi i\slash D {\bf
\Psi}^{(n)}+ \bar {\bf \Psi}^{(n)}i\slash D\psi \Big).
\end{array}
\ee Thus, along with $S$ all actions $S+\Delta S^{(n)}$ describe
the same physical theory. One might conclude that, if divergent
terms in the effective action are of the form (\ref{DS}), the
theory is renormalizable. The divergencies can be absorbed by the
SW redefinition: the new, redefined fields are the physical ones.
This, in some sense, generalizes the usual notion of
renormalizability and gives  an additional freedom to the theory.

A more detailed analysis of noncommutative electrodynamics with
fermions was done in~\cite{w} to first order in $\theta$. The
conclusion was that all propagators and vertices, with the
exception of the 4-fermion vertex, are renormalizable  if fermions
are massless. This result was confirmed to $\theta^2$-order for
the propagators in~\cite{mv}.

\initiate \section{Divergencies in $SU(2)$ and renormalization}

The divergencies of the one-loop effective action for $SU(2)$
gauge theory coupled to the massless fermions were calculated in
linear order in~\cite{mv1}.  The classical action of the theory is
(\ref{l0}-\ref{l1psi});  details about the quantization, gauge
fixing, ghosts etc. are explained in~\cite{mv1}. In the zero-th
order, the divergent part of the one-loop effective action is
given by \be \G_2 = \frac{1}{
    (4\pi)^2\e}\int d^4x\,\Big( \frac{10}{3}
F_{\m\n}^a F ^{\m\n a} +\frac{3i}{ 2}\bar\psi\slash D\psi\Big).
\ee Here and below the divergencies in the $n$-point functions are
expressed in the `covariantized' form. This means that for
example, the covariant 2-point function contains  parts of the
higher-point functions which are necessary to obtain the covariant
instead of the partial derivatives. The $\theta$-linear divergent
2-point function is \be \G_2^\prime = \frac{1}{
(4\pi)^2\e}\,\frac{i}{ 8}\,\theta^{\m\n}\e_{\m\n\r\s}\int
d^4x\,\bar\psi\g_5\g^\s D^2D^\r\psi ,\label{2gama}\ee whereas the
divergence in the 3-point function reads
\bea\label{3gama}\G^{\prime}_3&=&-\frac{1}{ (4\pi)^2\e}\,\frac
12\,\theta^{\m\n}\int d^4x\,\Big( -\frac{37i}{ 4}\,\bar\psi\g^\a
(F_{\n\a}D_\m + F_{\m\n}D_\a +F_{\a\m}D_\n)\psi\ncr
&-&\frac{6i}{4}\,\bar\psi\g^\a
F_{\m\n}D_\a\psi-\frac{3i}{4}\,\bar\psi\g^\a (D_\a
F_{\m\n})\psi\ncr &+& 2i\bar\psi\g_\m
F_{\n\a}D^\a\psi+i\bar\psi\g_\m (D^\a F_{\n\a})\psi\\ &+&\frac{5}{
8}\,\e_{\n\a\b\r} \bar\psi\g_5\g^\r (D_\m F^{\a\b})\psi-\frac
{1}{16}\,\e_{\m\n\a\b}\bar\psi\g_5\g^\s (D_\s F^{\a\b})\psi\ncr
&+&\frac {1}{8}\,\e_{\m\n\a\b}\big(2\bar\psi\g_5\g^\b
F^{\r\a}D_\r\psi+\bar\psi\g_5\g^\b(D_\r
F^{\r\a})\psi\big)\Big)\nonumber . \eea The divergent 4-fermion
vertex has the same form as in $U(1)$ theory: \be
\label{4gama}\G^\prime_{4\psi} =\frac{1}{
(4\pi)^2\e}\,\frac{9}{32}\,\theta^{\m\n}\e_{\m\n\r\s}\int
d^4x\,\bar\psi\g_5\g^\s\psi \, \bar\psi\g^\r\psi \ .\ee

Apparently, $\G_2^\prime$ and $\G_3^\prime $ are of the  form
given by (\ref{DS}). Therefore we could conclude that, as in
$U(1)$, the only obstacle to renormalization is the term
(\ref{4gama}). However, to prove this  we have to construct the
field redefinition  which removes the divergencies explicitly.
From the forms (\ref{2gama}-\ref{3gama})  we see that only the
fermionic field needs to be redefined. The most general
redefinition (\ref{red}) in the first order in $\theta$ is \be{\bf
\Psi}
^{(1)}=\theta^{\m\n}\Big(\kappa_1F_{\m\n}+i\kappa_2\s_{\m\r}F_\n{}^\r
+i\kappa_3\e_{\m\n\r\s}\g_5F^{\r\s}+\kappa_4\s_{\m\n}D^2+\kappa_5\s_{\r\m
}D_\n D^\r\Big)\psi ,\ee and has 5 free parameters,
$\kappa_1,\dots,\kappa_5$. The corresponding change in the
lagrangian (that is, its $\theta$-linear part) reads: \bea \Delta
\cl
^{(1)}&=&i\theta^{\a\b}\Big((\kappa_1-\frac12\,\kappa_5)(\bar\psi\g^\m
(D_\m F_{\a\b})\psi +2\bar\psi\g^\m F_{\a\b}D_\m\psi)\ncr
&-&\kappa_2(\bar\psi\g^\r (D_\a F_{\b\r})\psi +2\bar\psi\g^\r
F_{\b\r}D_\a\psi)\ncr &+&\kappa_2(\bar\psi\g_\a (D_\m F^{\
\m}_\b)\psi +2\bar\psi\g_\a F_{\b}^{\ \m}D_\m\psi)\ncr
&+&i\kappa_2\epsilon_{\m\a\r\s}\bar\psi\g_5\g^\s (D^\m F_{\b}^{\
\r})\psi - i\kappa_3\epsilon_{\a\b\r\s}\bar\psi\g_5\g^\m (D_\m
F_{\r\s})\psi \ncr &+&\kappa_4\epsilon_{\r\a\b\s}\bar\psi\g_5\g^\s
(2D^2D^\r\psi -2iF^{\r\m}D_\m\psi -i (D_\m F^{\m\r})\psi) \ncr
&+&2\kappa_4(\bar\psi\g_\b (D_\m F^{\ \m}_\a)\psi +2\bar\psi\g_\b
F_{\a\m}D^\m\psi)\Big).\eea Note  that, as  $\kappa_1$ and
$\kappa_5$ appear only in the combination  $\k_1 - \kappa_5/2$,
one of those two parameters is superfluous; we take $\kappa_5=0$.
After the shift $\psi\to\psi + {\bf \Psi^{(1)}}$ the renormalized
Lagrangian becomes \bea \label{27} && \cl+\cl_{ct} +\Delta
\cl^{(1)} = -\frac14 F_{\m\n}^aF^{\m\n
a}\Big(1+\frac{40g^2}{3(4\pi)^2\epsilon}\Big)\ncr &&+ i\bar\psi
\slash\pa\psi\Big(1-\frac{3g^2}{2(4\pi)^2\epsilon}\Big)
+g\m^{\frac\epsilon 2}\bar\psi
A^a_\m\g^\m\frac{\s^a}{2}\psi\Big(1-\frac{3g^2}{2(4\pi)^2\epsilon}\Big)\ncr
&&+ig \m^{\frac\epsilon 2}\theta^{\a\b} \bar\psi\g^\m\big(
F_{\b\m}D_\a + F_{\a\b}D_\m +F_{\m\a}D_\b\big)\psi \,\Big(-\frac
14-\kappa _2 - \frac{37 g^2}{8(4\pi)^2\epsilon}\Big) \ncr && +
ig\m^{\frac\epsilon 2}\theta^{\a\b} \big( \bar\psi\g^\m (D_\m
F_{\a\b})\psi +2\bar\psi\g^\m F_{\a\b}D_\m\psi \big) \Big(\kappa_1
+\frac {\kappa_2}{2} -\frac{3g^2}{8(4\pi)^2\epsilon} \Big)\ncr &&+
ig\m^{\frac\epsilon 2}\theta^{\a\b} \big(\bar\psi\g_\a (D^\m
F_{\b\m})\psi+ 2\bar\psi\g_\a F_{\b\m}D^\m\psi\big)
\Big(\kappa_2-2\k_4+\frac{g^2}{2(4\pi)^2\epsilon}\Big)\ncr &&
-g\m^{\frac\epsilon 2}\theta^{\a\b}\epsilon_{\b\m\n\r}
\bar\psi\g_5\g^\r (D_\a F^{\m\n})\psi\, \Big(\frac {\kappa_2}{2}-
\frac{5g^2}{16(4\pi)^2\epsilon}\Big)\\ && +g\m^{\frac\epsilon
2}\theta^{\a\b}\epsilon_{\a\b\r\s} \bar\psi\g_5\g^\m (D_\m
F^{\r\s})\psi
\,\Big(\kappa_3-\frac{g^2}{32(4\pi)^2\epsilon}\Big)\ncr &&
+g\m^{\frac\epsilon 2}\theta^{\a\b}\epsilon_{\a\b\r\s}\big(
\bar\psi\g_5\g^\s (D_\m F^{\r\m})\psi+2\bar\psi\g_5\g^\s
F^{\r\m}D_\m\psi\big)
\Big(\kappa_4-\frac{g^2}{16(4\pi)^2\epsilon}\Big)\ncr &&
-g\m^{\frac\epsilon 2}\theta^{\a\b}\epsilon_{\a\b\r\s}
\bar\psi\g_5\g^\s D^2D^\r\psi\,
\Big(2\kappa_4-\frac{g^2}{8(4\pi)^2\epsilon}\Big) .\nonumber
\label{55} \eea If we choose  $$
\k_1=\frac{g}{16(4\pi)^2\epsilon},\quad
\k_2=\frac{5g}{8(4\pi)^2\epsilon},\quad
\k_3=\frac{g}{32(4\pi)^2\epsilon},\quad
\k_4=\frac{g}{16(4\pi)^2\epsilon} ,$$ the expression (\ref{27})
reduces to \bea \label{28} && \cl+\cl_{ct} +\Delta \cl^{(1)} =
-\frac14 F_{\m\n}^aF^{\m\n
a}\Big(1+\frac{40g^2}{3(4\pi)^2\epsilon}\Big)\ncr &&+ i\bar\psi
\slash\pa\psi\,\Big(1-\frac{3g^2}{2(4\pi)^2\epsilon}\Big)
+g\m^{\frac\epsilon 2}\bar\psi
A^a_\m\g^\m\frac{\s^a}{2}\psi\,\Big(1-\frac{3g^2}{2(4\pi)^2\epsilon}\Big)\\
&&-\frac i4\, g \m^{\frac\epsilon 2}\theta^{\a\b}
\bar\psi\g^\m\big( F_{\b\m}D_\a + F_{\a\b}D_\m
+F_{\m\a}D_\b\big)\psi \,\Big(1 +\frac{21
g^2}{(4\pi)^2\epsilon}\Big) \ncr &&+ ig\m^{\frac\epsilon
2}\theta^{\a\b} \big(\bar\psi\g_\a (D^\m F_{\b\m})\psi+
2\bar\psi\g_\a F_{\b\m}D^\m\psi\big) \frac{g^2}{(4\pi)^2\epsilon}
\ .\nonumber \eea Now, one would introduce the bare fields and the
couplings as \bea
&&\psi_0=\sqrt{Z_2}\psi=\sqrt{1-\frac{3g^2}{2(4\pi)^2\epsilon}}\,\psi
,\ncr && A^\m_0=\sqrt{Z_3}A^\m
=\sqrt{1+\frac{40g^2}{3(4\pi)^2\epsilon}}\,A^\m , \\
&&g_0=g\m^{\frac{\epsilon}{2}}Z_3^{-1/2}Z_2^{-1}\Big(1-\frac{3g^2}{2(4\pi)^2\epsilon}
\Big)
=g\m^{\frac{\epsilon}{2}}\Big(1-\frac{20g^2}{3(4\pi)^2\epsilon}\Big)\ncr
&& \theta^{\a\b}_0 =\theta ^{\a\b} {Z_2}^{-1}{Z_3}^{-\frac12}
\Big( 1+\frac {21g^2}{(4\pi)^2\epsilon
}\Big)\Big(1-\frac{20g^2}{3(4\pi)^2\epsilon}\Big)^{-1}=
\theta^{\a\b}\Big(1+\frac{45g^2}{2(4\pi)^2\epsilon} \Big)
,\nonumber \eea were the last term in (\ref{28}) absent. Then the
renormalized Lagrangian would be \bea && \cl+\cl_{ct} +\Delta
\cl^{(1)} = -\frac14 F_{\m\n 0}^aF^{\m\n a}_0 + i\bar\psi_0
\slash\pa\psi_0 +g_0\bar\psi_0 \g_\m A^{\m
a}_0\frac{\s^a}{2}\psi_0\ncr &&-\frac i4 g_0\theta^{\a\b}_0
\bar\psi_0\big( F_{\b\m 0}D_\a + F_{\a\b 0}D_\m +F_{\m\a
0}D_\b\big)\psi_0
\\&&+ ig\m^{\frac\epsilon
2}\theta^{\a\b} \big(\bar\psi\g_\a (D^\m F_{\b\m})\psi+
2\bar\psi\g_\a F_{\b\m}D^\m\psi\big) \frac{g^2}{(4\pi)^2\epsilon}
. \nonumber \eea
 Unfortunately, the
last term, which is divergent, remains: it cannot be absorbed by
SW the redefinition or by the renormalizion procedure.

\initiate \section{Conclusion} As we saw, the SW redefinition does
not possess enough free parameters to absorb  the divergencies of
the 3-point functions. Thus we come to the conclusion: the
$\theta$-expanded $SU(2)$ theory is not renormalizable even for
massless fermions. Apart from the 4-fermion vertex, 3-vertices
also cannot be renormalized. This result differs from the
corresponding one for  noncommutative $U(1)$, \cite{w}. Comparing
the two, we may add that our technique is slightly simpler: we
introduce the SW redefinition after the quantization and not
already in the classical action. This however does not change the
reasoning essentially.

The negative result about renormalizability which we presented is
conclusive, as clearly it cannot be altered in  higher orders in
$\theta$. The presence of fermions prevents renormalizability,
except of course in the supersymmetric theories. The behavior of
the pure gauge theories on the other hand is still not fully clear
and deserves further investigation.

\end{document}